# Reconstruction of Protein-Protein Interaction Pathways by Mining Subject-Verb-Objects Intermediates

Maurice HT Ling[1,2], Christophe Lefevre[3],
Kevin R. Nicholas[2], Feng Lin[1]

[1] BioInformatics Research Centre, Nanyang Technological University, Singapore
[2] CRC for Innovative Dairy Products, Department of Zoology,
The University of Melbourne, Australia
[3] Victorian Bioinformatics Consortium, Monash University, Australia
mauriceling@acm.org, k.nicholas@zoology.unimelb.edu.au,
Chris.Lefevre@med.monash.edu.au, ASFLIN@ntu.edu.sg

**Abstract.** The exponential increase in publication rate of new articles is limiting access of researchers to relevant literature. This has prompted the use of text mining tools to extract key biological information. Previous studies have reported extensive modification of existing generic text processors to process biological text. However, this requirement for modification had not been examined. In this study, we have constructed Muscorian, using MontyLingua, a generic text processor. It uses a two-layered generalization-specialization paradigm previously proposed where text was generically processed to a suitable intermediate format before domain-specific data extraction techniques are applied at the specialization layer. Evaluation using a corpus and experts indicated 86-90% precision and approximately 30% recall in extracting protein-protein interactions, which was comparable to previous studies using either specialized biological text processing tools or modified existing tools. Our study had also demonstrated the flexibility of the two-layered generalization-specialization paradigm by using the same generalization layer for two specialized information extraction tasks.

**Keywords:** biomedical literature analysis, protein-protein interaction, montylingua

## 1 Introduction

PubMed currently indexes more than 16 million papers with about one million papers and 1.2 million added in the years 2005 and 2006 respectively. A simple keyword search in PubMed showed that nearly 900 thousand papers on mouse and more than 1.3 million papers on rat research had been indexed in PubMed to date, and in the last four years, more than 150 thousand papers have been published on each of mouse and rat research. This trend of increased volume of research papers indexed in PubMed over the last 10 years makes it difficult for researchers to maintain an active and productive assessment of relevant literature. Information extraction (IE) has been used as a tool to analyze biological text to derive assertions on specific biological domains [30], such as protein phosphorylation [19] or entity interactions [1].

A number of IE tools used for mining information from biological text can be classified according to their capacity for general application or tools that considers biological text as specialized text requiring domain-specific tools to process them. This has led to the development of specialized part-of-speech (POS) tag sets (such as SPECIALIST [28]), POS taggers (such as MedPost [33]), ontologies [11], text processors (such as MedLEE [15]), and full IE systems, such as GENIES [16], MedScan [29], MeKE [4], Arizona Relation Parser [10], and GIS [5]. On the other hand, an alternative approach assumes that biological text are not specialized enough to warrant re-development of tools but adaptation of existing or generic tools will suffice. To this end, BioRAT [12] had modified GATE [8], MedTAKMI [36] had modified TAKMI [27], originally used in call centres, Santos [31] had used Link grammar parser [32].

Although both systems demonstrated similar performance, either developing these systems or modifying existing systems were time consuming [20]. Although work by Grover [17] suggested that native generic tools may be used for biological text, a recent review had highlighted successful uses of a generic text processing system, MontyLingua [14, 23], for a number of purposes [22]. For example, MontyLingua has been used to process published economics papers for concept extraction [35]. The need to modify generic text processors had not been formally examined and the question of whether an un-modified, generic text processor can be used in biological text analysis with comparable performance, remains to be assessed.

In this study, we evaluated a native, generic text processing system, MontyLingua [23], in a two-layered generalization-specialization architecture [29] where the generalization layer processes biological text into an intermediate knowledge representation for the specialization layer to extract genic or entity-entity interactions. This system demonstrated 86.1% precision using Learning Logic in Languages 2005 evaluation data [9], 88.1% and 90.7% precisions in extracting protein-protein binding and activation interactions respectively. Our results were comparable to previous work which modified generic text processing systems which reported precision ranging from 53% [24] to 84% [5], suggesting this modification may not improve the efficiency of information retrieval.

## 2  System Description

We have developed a biological text mining system, known as Muscorian, for mining protein-protein inter-relationships in the form of subject-relation-object (for example, protein X bind protein Y) assertions. Muscorian is implemented as a 3-module sequential system of entity normalization, text analysis, and protein-protein binding finding, as shown in Figure 1. It is available for academic and non-profit users through http://ib-dwb.sf.net/Muscorian.html.

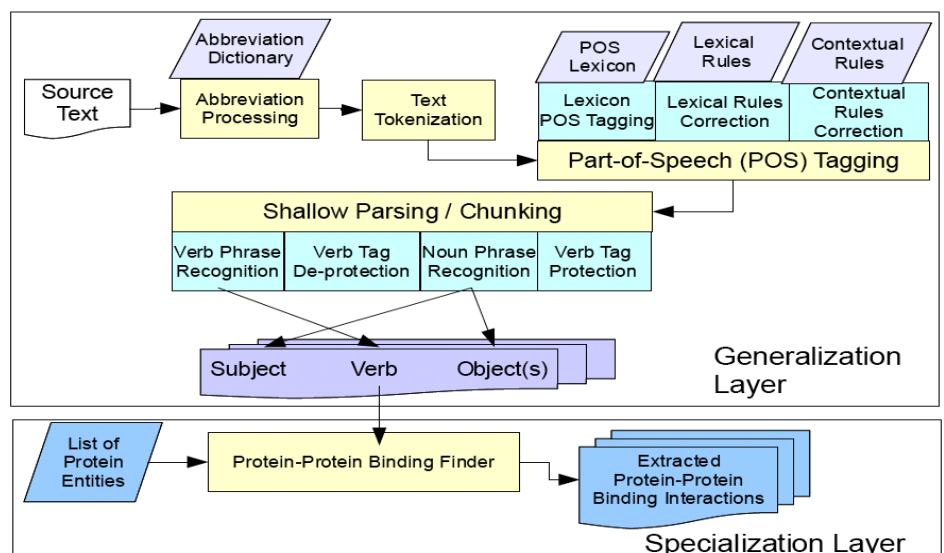

Fig 1. Schematic Diagram Illustrating the Operations of Muscorian

### 2.1 Entity Normalization

Entity normalization is the substitution of the long form of either a biological or chemical term with its abbreviated form. This is essential to correct part-of-speech tagging errors which are common in biological text due to multi-worded nouns. For example, the protein name "phosphatase and tensin homolog deleted on chromosome 10" has to be recognized as a single noun and not a phrase. In this study, we attempt to mine protein-protein interactions and consolidate this knowledge to produce a map. Therefore, the naming convention of the protein entities must be standardized to allow for matching. However, this is not the case for biological text and synonymous protein names exist for virtually every protein. For example, "MAP kinase kinase", "MAPKK", "MEK" and "MAPK/Erk kinase" referred to the same protein. Both of these problems could be either resolved or minimized by reducing multi-worded nouns into their abbreviated forms.

A dictionary-based approach was used for entity normalization to a high level of accuracy and consistency. The dictionary was assembled as follows: firstly, a set of 25000 abstracts from PubMed was used to interrogate Stanford University's BioNLP server [3] to obtain a list of long forms with its abbreviations and a calculated score. Secondly, only results with the score of more than 0.88 were retained as it is an inflection point of ROC graph [3], which is a good balance between obtaining the most information while reducing curation efforts. Lastly, the set of long form and its abbreviations was manually curated with the help of domain experts.

The domain experts curated dictionary of long forms and its abbreviated term was used to construct a regular expression engine for the process of recognition of the long form of a biological or chemical term and substituting it with its corresponding abbreviated form.

## 2.2 Text Analysis

Entity normalized abstracts were then analyzed textually by an un-modified text processing engine, MontyLingua [14], where they were tokenized, part-of-speech tagged, chunked, stemmed and processed into a set of assertions in the form of 3-element subject-verb-object(s) (SVO) tuple, or more generally, subject-relation-object(s) tuple. Therefore, a sequential pattern of words which formed an abstract was transformed through a series of pattern recognition into a set of structurally-definable assertions.

Before part-of-speech tagging is possible, an abstract made up of one or more sentences had to be separated into individual sentences. This is done by regular expression recognition of sentence delimiters, such as full-stop, ellipse, exclamation mark and question mark, at the end of a word (regular expression: ([?!]+|[.][.]+)$) with an exception of acronyms. Acronyms, which are commonly represented with a full-stop, for example "Dr.", are not denoted as the end of a sentence and were generally prevented by an enumeration of common acronyms.

Individual sentences were then separated into constituent words and punctuations by a process known as tokenization. Tokenization, which is essential to atomize a sentence into atomic syntactic building blocks, is generally a simple process of splitting of an English sentence in words using whitespaces in the sentence, resulting in a list of tokens (words). However, there were three problems which were corrected by examining each token. Firstly, punctuations are crucial in understand a written English sentence, but typographically a punctuation is usually joined to the presiding word. Hence, punctuation separation from the presiding word is necessary. However, it resulted in incorrect tokenization with respect to acronyms and decimal numbers. For example, "... an appt. for ..." will be tokenized to "... an appt . for ..." and "$4.20" will be "$ 4 . 20". This problem was prevented by pre-defining acronyms and using regular expressions, such as "^[$][0-9]{1,3}[.][0-9][0-9](?[.]?)$". Lastly, common abbreviated words, such as "don't", were expanded into two tokens of "do" and "n't". Despite the above error correction measures, certain text such as mathematical equations, which might be used to describe enzyme kinetics in biological text, will not be tokenized correctly. In spite of this limitation, the described tokenization scheme is still appropriate as extraction of enzyme kinetics or mathematical representations are not the aims of this study.

Each of the tokens (words and punctuations) in a tokenized sentence is then tagged using Penn TreeBank Tag Set [25] by a Brill Tagger, trained on Wall Street Journal and Brown corpora, which operates in two phases. Using a lexicon, containing the likely tag for each word, each word is tagged. This is followed by a phase of correction using lexical and contextual rules, which were learnt using training with a tagged corpora, in this case, Wall Street Journal and Brown corpora. Lexical rules uses a combination of preceding tag and prefix or suffix of the token (word) in question. For example, the rule "NN ing fhassuf 3 VBG" defines that if the current token is tagged as a noun (NN) and has a 3-character suffix of "ing", then the tag should be a verb (VBG). On the other hand, contextual rules uses only the preceding or proceeding tags and hence, must be applied after lexical rules for effectiveness. The contextual rule "RB JJ NEXTTAG NN" defines that an abverbial tag (RB) should be changed to an adjective (JJ) if the next token was tagged as a noun (NN). A table of Penn Treebank Tag Set [25] without punctuation tags is given in Table 1.

| Tag | Description | Tag | Description |
|---|---|---|---|
| CC | Coordinating conjunction | PRP$ | Possessive pronoun |
| CD | Cardinal number | RB | Adverb |
| DT | Determinant | RBR | Adverb, comparative |
| EX | Existential *there* | RBS | Adverb, superlative |
| FW | Foreign word | RP | Particle |
| IN | Preposition or subordinating conjunction | SYM | Symbol |
| JJ | Adjective | TO | to |
| JJR | Adjective, comparative | UH | Interjection |
| JJS | Adjective, superlative | VB | Verb, base form |
| LS | List item marker | VBD | Verb, past tense |
| MD | Modal | VBN | Verb, past participle |
| NN | Noun, singular or mass | VBG | Verb, gerund or present participle |
| NNS | Noun, plural | VBP | Verb, non-3$^{rd}$ person singular present |
| NNP | Proper noun, singular | VBZ | Verb, 3$^{rd}$ person singular present |
| NNPS | Proper noun, plural | WDT | Wh-determiner |
| PDT | Predeterminer | WP | Wh-pronoun |
| POS | Possessive ending | WP$ | Possessive wh-pronoun |
| PRP | Personal pronoun | WRB | Wh-adverb |

Table 1. Penn Treebank Tag Set without Punctuation Tags (Adapted from [25])

By tagging, the complexity of an English sentence (ie, the number of ways an English sentence can be grammatically constructed with virtually unlimited words and unlimited ideas) was collapsed into a sequence of part-of-speech tags, in this case, Penn TreeBank Tag Set [25], with only about 40 tags. Therefore, tagging reduced the large number of English words to about 40 "words" or tags.

Generally, an English sentence is composed of a noun phrase, a verb, and a verb phase, where the verb phrase may be reduced into more noun phrases, verbs, and verb phrases. More precisely, the English language is an example of subject-verb-object typology structure, which accounts for 75% of all languages in the world [7]. This concept of English sentence structure is used to process a tagged sentence into higher-order structures of phrases by a process of chunking, which is a precursor to the extraction of semantic relationships of nouns into SVO structure. Using only the sequence of tags, chunking was performed as a recursive 4-step process: protecting verbs, recognition of noun phrases, unprotecting verbs and recognition of verb phrases. Firstly, verb tags (VBD, VBG and VBN) were protected by suffixing the tags. The main purpose was to prevent interference in recognizing noun phrases. Secondly, noun phrases were recognized by the following regular expression pattern of tags:

```
((((PDT )?(DT |PRP[$] |WDT |WP[$] )(VBG |VBD |VBN |JJ |
JJR |JJS |, |CC |NN |NNS |NNP |NNPS |CD )*(NN |NNS |NNP
|NNPS |CD )+)|((PDT )?(JJ |JJR |JJS |, |CC |NN |NNS |
NNP |NNPS |CD )*(NN |NNS |NNP |NNPS |CD )+)|EX |PRP |WP
|WDT )POS )?(((PDT )?(DT |PRP[$] |WDT |WP[$] )(VBG |VBD
|VBN |JJ |JJR |JJS |, |CC |NN |NNS |NNP |NNPS |CD )*(NN
|NNS |NNP |NNPS |CD )+)|((PDT )?(JJ |JJR |JJS |, |CC |
NN |NNS |NNP |NNPS |CD )*(NN |NNS |NNP |NNPS |CD )+)|EX
|PRP |WP |WDT )
```

Thirdly, the protected verb tags in the first step were de-protected by removing the suffix appended onto the tags. Lastly, verb phrases were recognized by the following regular expression:

```
(RB |RBR |RBS |WRB )*(MD )?(RB |RBR |RBS |WRB )*(VB |
VBD |VBG |VBN |VBP |VBZ )(VB |VBD |VBG |VBN |VBP |VBZ |
RB |RBR |RBS |WRB )*(RP )?(TO (RB )*(VB |VBN )(RP )?)?
```

After chunking, each word (token) was stemmed into its root or infinite form. Firstly, each word was matched against a set of rules for specific stemming. For example, the rule "dehydrogenised verb dehydrogenate" defines that if the word "dehydrogenised" was tagged as a verb (VBD, VBG and VBN tags), it would be stemmed into "dehydrogenate". Similarly, the words "binds", "binding" and "bounded" were stemmed to "bind". Secondly, irregular words which could not be stemmed by removal of prefixes and suffixes, such as "calves" and "cervices", were stemmed by a pre-defined dictionary. Lastly, stemming was done by simple removal of prefixes or suffixes from the word based on a list of common prefixes or suffixes. For example, "regards" and "regarding" were both stemmed into "regard".

Given the general nature of an English sentence is an aggregation of noun phrase, a verb, and a verb phase, where the verb phrase may be reduced into more noun phrases, verbs, and verb phrases, each verb phrase may be taken as a sentence by itself. This allowed for recursive processing of a chunked-stemmed sentence into SVO(s) by a 3-step process. Firstly, the first terminal noun phrase, delimited by "(NX" and "NX)" was taken as the subject noun. Secondly, proceeding from the first terminal noun phrase, the first terminal verb would be taken as the verb in the SVO. Lastly, the rest of the phrase was scanned for terminal noun phrases and would be taken as the object(s). The recursive nature of SVO extraction also meant that the subject, verb, and object(s) will be contiguous, which had been demonstrated to have better precision than non-contiguous SVOs [26].

### 2.3 Protein-Protein Binding Finding

The protein-protein binding finder module is a data miner for protein-protein binding interaction assertions from the entire set of subject-relation-object (SVO) assertions from the text analysis process using apriori knowledge. That is, the set of proteins of interest must be known, in contrast to an attempt to uncover new protein entities, and their binding relationships with other protein entities, that were not known to the researcher.

Protein-protein binding assertions were extracted in a three step process. Firstly, a set of SVOs was isolated by the presence of the term "bind" in the verb clause resulting in a set of "bind-SVOs" assertions. Non-infinite forms of "bind" (such as, "binding" and "binds") were not used as verbs were stemmed into their infinite forms during text processing. Secondly, the set of bind-SVOs were further characterized for the presence of protein entities in both subject and object clauses by comparing with the desired list of protein entities. A pairwise isolation of bind-SVOs for protein entities resulted in a set of bind-SVOs, "entity-bind-SVOs", containing SVOs describing binding relationship between the protein entities. Lastly, entity-bind-SVOs were cleaned so that the subject and object clauses only contains protein entities. For example, "MAPK in the cytoplasm" in the object clause will be reduced to just the entity name "MAPK", the full subject and object clauses could be used in other information extraction tasks, such as determining protein localization, but is not explored in this study. This step is required to allow for the construction of network graphs, such as using Graphviz, without reference to the list of protein names during construction. Given that protein_entities is the list of desired proteins, table SVO contains the SVO output from MontyLingua and table entity_bind_SVO contains the isolated and cleaned SVOs, the pseudocode for Protein-Protein Binding Finding module is given as:

```
for subject_protein in protein_entities_{1 to n}
    for object_protein in protein_entities_{1 to n}
        insert (pmid, subject_protein, object_protein) into entity_bind_SVO
            from select pmid
            from (select * from SVO where verb = 'bind')
            where subject is containing subject_protein
            and object is containing object_protein
```

### 3 Experimental Results

Four experiments were carried out to evaluate the performance of Muscorian and demonstrate the flexibility of the two-layered generalization-specialization approach in constructing systems that could be readily be adapted to related problems. The results are summarized in Table 2.

|  | *LLL05 Directional* | *LLL05 Un-directional* | *Protein-Protein Binding* | *Protein-Protein Activation* |
|---|---|---|---|---|
| Precision | 55.8% | 86.1% | 88.1% | 90.7% |
| Recall | 19.8% | 30.7% | Not measured | Not measured |

Table 2. Summary of the Experimental Results Comparing the Precision and Recall Measures.

### 3.1 Benchmarking Muscorian Performance

The performance of Muscorian, in terms of precision and recall, could only be evaluated using a defined data set with known results. For such purpose, the data set for Learning Languages in Logic 2005 (LLL05) [9] was used to benchmark Muscorian on genic interactions, which is a superset of protein-protein binding interactions. LLL05 had defined a genic interaction as an interaction between 2 entities (agent and target) but the nature of interaction was not considered under the challenge task. LLL05 provided a list of protein entities found in the data set, which was used to filter subject-relation-object assertions from text analysis (MontyLingua) output where both subject and object contained protein entities in the given list. The filtered list of assertions was evaluated for precision and recall, which was found to be 55.6% and 19.8% respectively.

LLL05 required that the agent and target (subject and object) to be in the correct direction, making it a vector quality. However, this requirement was not biologically significant to protein-protein binding interactions, which is scalar. For example, "X binds to Y" and "Y binds to X" have no biological difference. Hence, this requirement of directionality was eliminated and the precision and recall was 86.1% and 30.7% respectively.

### 3.2 Verifying Protein-Protein Binding Interactions

Precision of Muscorian for mining protein-protein binding interactions from published abstracts was evaluated by manual verification of a sample of assertions (n=135) yielded by the protein-protein binding finder module against the original abstracts. Each of the sampled assertions was assumed to be atomic, in the form of "X binds Y". In cases where there were more than one target, such as "X binds Y and Z", they would be reduced to atomic assertions. In this case, "X binds Y and Z" would be reduced to 2 assertions, "X bind Y" and "X bind Z". These were then checked with the original abstract, traceable by the PubMed IDs, and precision was measured as the ratio of the number of correct assertions to the number of sampled atomic assertions (which is 135). A 95% confidence interval was estimated by bootstrapping (re-sampling with replacement) [13] of the manual verification results. Our results suggested a precision of 88.1%, with a 95% confidence interval between 82.4% to 93.7%.

An IE trial was performed using the Protein-Protein Binding Finding module to search for the binding partners of CREB and insulin receptor and a sample network diagram of the results are shown in Figure 2 and 3 respectively.

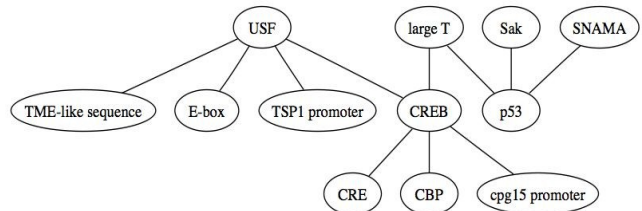

Fig 2. Preliminary Protein Binding Network of CREB

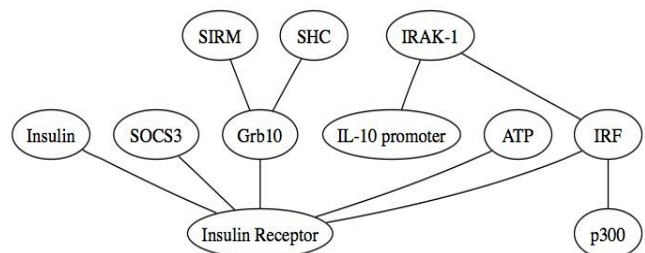

Fig 3. Preliminary Protein Binding Network of Insulin Receptor

### 3.3 Large Scale Mining of Protein-Protein Binding Interactions

A large scale mining of protein-protein binding interactions was carried out using all of the PubMed abstracts on mouse (about 860000 abstracts), which were obtained using "mouse" as the keyword for searches, with a predefined set of about 3500 abbreviated protein entities as the list of proteins of interest (available from http://cvs.sourceforge.net/viewcvs.py/ib-dwb/muscorian-data/protein_accession.csv?rev=1.2&view=markup). In this experiment, the primary aim was to apply Muscorian to large data set and the secondary aim was to look for multiple occurrences of the same interactions as multiple occurrences might greatly improve precision confidence.

For example, given our lower confidence estimate that the precision of Muscorian with respect to mining protein-protein binding interactions is 82%, which means that every binding assertion has an 18% likelihood of not having a corresponding representation in the published abstracts. However, if 2 abstracts yielded the same binding assertion, the probability of both being wrong was reduced to 3.2% ($0.18^2$), and the corresponding probability that at least one of the 2 assertions was correctly represented was 96.8% ($1-0.18^2$). The more times the same assertion was extracted from multiple sources text (abstracts), the higher the possibility that the mined interaction was represented at least once in the set of abstracts. For example, if 5 abstracts yielded the same assertion, the possibility that at least one of the 5 assertions was correctly represented would be 99.98% ($1-0.18^5$).

Our experiment mined a total of 9803 unique protein-protein binding interactions, of which 7049 binding interactions were from one abstract (P=82%), 1297 binding interactions were from two abstracts (P=96.8%), 516 binding interactions were from three abstracts (P=99.4%), 235 binding interactions were from four abstracts

(P=99.9%), 164 binding interactions were from five abstracts (P=99.98%), 105 binding interactions were from six abstracts (P=99.997%), 69 binding interactions were from seven abstracts (P=99.9993%), 398 binding interactions were from more than seven abstracts (P>99.9993%).

### 3.4 Pilot Study - Protein-Protein Activation Interactions

In order to demonstrate the adaptability of our proposed two-layered model, a small pilot study for mining protein-protein activation interactions was carried out. For this study, the protein-protein binding finder module, the data mining module for mining protein-protein binding interaction, was replaced with a protein-protein activation finder module.

The protein-protein activation finder was semantically similar to the original protein-protein binding finder module as described in Section 3.3 previously. The only difference was that raw assertion output from MontyLingua was filtered for activation-related assertions, instead of binding-related assertions, before analysis for the presence of protein names in both subject and object nouns from a pre-defined list of proteins of interest. For example, by modifying the Protein-Protein Binding Finding module to look for the verb 'activate' instead of 'bind', it can then be used for mining protein-protein activation interactions. A trial was done for insulin activation and a subgraph is illustrated in Figure 4 below.

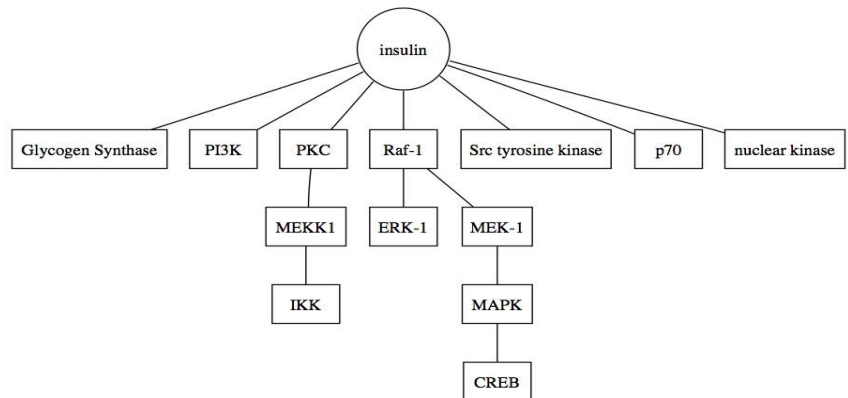

Fig 4. Preliminary Protein Activation Network of Insulin

The precision measure of Muscorian for mining protein-protein activation interactions was calculated using identical means as described for protein-protein binding interactions. Using a sample of 85 atomic assertions, the precision of Muscorian for mining protein-protein activation interactions was estimated to be 90.7%, with a 95% confidence interval of precision between 84.7% to 96.4% by bootstrapping [13].

## 4  Discussion

New research articles in gene expression regulation networks, protein-protein interactions and protein docking are emerging at a rate faster than what most biologists can manage to extract the data and generate working pathways. Information extraction technologies have been successfully used to process research text and automate fact extraction [1]. Previous studies in biological text mining have developed specialized text processing tools and adapted generic tools to relatively good performance of more than 80% in precision [5, 11, 20, 31]. However, either specialized tool development or modifying existing tools often require much effort [20]. The need to modify existing tools has not been formally tested and the possibility of using an un-modified generic text processor for biological text for the purpose of extracting protein-protein interaction remains unresolved. Using a two-layered approach [29] of generalizing biological text into a structured intermediate form, followed by specialized data mining, we present Muscorian, which uses MontyLingua natively in the generalized layer, as a tool for extracting either protein-protein or genic interactions from about 860000 published biological abstracts.

Benchmarking Muscorian against LLL05, a tested data set, demonstrated a precision of 55.6%, which is about 5% higher than that reported in the conference and a recall of 19.7% is similar to that reported by other participants of LLL05 [9]. This may be due to the emphasis of LLL05 on F-measure, which is the harmonic mean of precision and recall, rather than putting more emphasis on precision. Nevertheless, this also suggested that Muscorian is able to perform text analysis for the purpose of extracting genic interactions effectively, which is comparable to specialized systems reported in LLL05. In addition, directionality of genic interactions was not a concern for protein-protein binding interactions as binding interaction is scalar rather than vector. By eliminating directionality of genic interactions, the precision and recall of Muscorian was 86.1% and 30.7% respectively. This suggested that Muscorian is a suitable tool for mining quality genic interactions from biological text compared to other tools reported in LLL05 [9].

Our results on protein-protein binding and activation interactions show the insulin receptor binds to IL-10 promoter through IRF and IRAK-1, which is an important insulin receptor signalling pathway. In addition, our data shows insulin activates CREB via Raf-1, MEK-1 and MAPK, which is consistent with the MAP kinase pathway. Combining these data (Figures 2 and 4) indicated that insulin activates CREB via MAP kinase pathway, and CREB binds to cpg15 promoter in the nucleus. A simple keyword search on PubMed, using the term "cpg15 and insulin" (done on 30$^{th}$ of April, 2007), did not yield any results, suggesting that the effects of insulin on cpg15, also known as neuritin [2], had not been studied thoroughly. This might also suggest limited knowledge shared between insulin investigators and cpg15 investigators as suggested by Don Swanson in his classical paper describing the links between fish oil and Raynaud's syndrome [34]. Neuritin is a relatively new research area with less than 20 papers published (as of 30$^{th}$ of April, 2007) and had been implicated as a lead for neural network re-establishment [18], suggesting potential collaborations between endocrinologists and neurologists.

Our experiments in extracting two different forms of relations demonstrated that despite using specialized dictionaries in the generalized layer, it is still general to the extend that specific application (the type of relationships to extract) was not built into the generalized layer.

At the same time, these 2 experiments also illustrated the relative ease in re-targeting the system for extracting another form of relationship by modifying the specialized layer. The Protein-Protein Activation Finder module is a slight modification of the original Protein-Protein Binding Finder module where the original SQL statement that selects 'bind'-related SVOs from total SVOs, "*select \* from SVO where verb = 'bind'*", was changed to "*select \* from SVO where verb = 'activate'*" to select for 'activation'-related SVOs from total SVOs. Hence, it is plausible that similar changes may suffice for extracting other relationships, such as 'inhibition'. This relative ease of re-targeting the system for extracting other relationships also demonstrated the robustness of the generalization layer, as implied by Novichkova et. al. [29] – "*the adaptability of the system to related problems other than the problem the system was designed for*".

Given large numbers of published abstracts, the performance of Muscorian on precision was comparable with published values of BioRAT (58.7%) [12], GIS (84%) [5], Cooper and Kershenbaum (74%) [6] and CONAN (53%) [24] while Muscorian's recall was comparable with published values of Arizona Relations Parser (35%) [10] and Daraselia et. al. (21%) [11]. Poor precision was considered unacceptable because incorrect information is more detrimental than missing information (1 - recall) when protein-protein binding interactions were used to support other biological analyses. Muscorian's mediocre recall of 30% (from LLL05 test set evaluation) could be supplemented by the fact that the same interaction could be mentioned or described by multiple abstracts; thus, the actual recall when tested on a large corpus may be higher. For example, 30% recall essentially means a loss of 70% of the information; however, if the same information (in this case, protein interactions) were mentioned in 3 or more abstracts, there is still a reasonable chance to believe that information from at least 1 of the 3 or more abstracts will be extracted. This is supported by our results indicating that almost 30% (2754 of 9803) of binding interactions were extracted from more than one abstract.

Multiple isolation of 2754 binding interactions enabled a higher confidence that these interactions were correctly extracted with reference to the source literature. Based on this analysis, 2754 binding interactions could be assigned higher confidence based on their occurrences [21], in this case more than 95% chance of being correct based on literature. In addition, the number of multiple interaction occurrence varies inversely with the number of abstracts these interactions were found in is in line with expectation. Although this line of argument is based on the assumption that the appearance of protein names across abstracts were independent, it can be reasonably held as this study uses abstracts rather than full text – abstracts tends to describe what main results of the particular article while the introduction of a full text article tends to be a brief background review of the field. Hence, independence of protein names can be better assumed in abstracts than in full text articles.

An evaluation of a sample of atomic assertions (interactions) of binding and activation interactions between entities was performed by domain experts comparing the assertions with their source abstracts. Both approaches gave similar precision measures and are consistent with the evaluation using LLL05 test set. The ANOVA test demonstrated that there was no significant differences between these three precision measures. Taken together, these evaluations strongly suggested that Muscorian performed with precisions between 86-90% for genic (gene-protein and protein-protein) interactions, which was similar to that reported by studies either modifying existing tools [31] or developing specialized tools [11]. This suggested that MontyLingua could be used natively (un-modified), with good precision, to process biological text into structured subject-verb-objects tuples which could be mined for protein interactions.

**Acknowledgments.** We wish to thank Prof. I-Fang Chung, Institute of Biomedical Informatics, National Yang Ming University, Taiwan, for his comments on improving the initial drafts. This work is sponsored by the CRC for Innovative Dairy Products, Australia, and Postgraduate Overseas Research Experience Scholarship, The University of Melbourne, Australia.